\documentclass[aps,prl,twocolumn,showpacs,floatfix,nofootinbib,superscriptaddress]{revtex4-1}
\usepackage{amsmath}
\usepackage{amssymb}
\usepackage{graphicx}
\usepackage{hyperref}
\usepackage{color}
\usepackage{mathtools}
\usepackage{times}

\newcommand{\up}[1]{{\rm #1}}
\newcommand{\Tobs}{T^\up{obs}}
\newcommand{\Tbobs}{\left\langle T\right\rangle^\up{obs}}
\newcommand{\Dobs}{d^\up{obs}}
\newcommand{\aobs}{a^\up{obs}}
\newcommand{\Tbar}{\bar T}
\newcommand{\bdv}[1]{{\bf #1}}
\newcommand{\beeq}{\begin{equation}}
\newcommand{\eneq}{\end{equation}}
\newcommand{\bear}{\begin{eqnarray}}
\newcommand{\enar}{\end{eqnarray}}
\newcommand{\nn}{\nonumber}
\newcommand{\pa}{\partial}
\newcommand{\nhat}{\hat n}
\newcommand{\Tr}{\up{Tr}}
\newcommand{\ga}{\gamma}
\newcommand{\de}{\delta}

\begin{document}

\title{$\Tbar$: A New Cosmological Parameter?}

\author{Jaiyul Yoo}
\email{jyoo@physik.uzh.ch}
\affiliation{Center for Theoretical Astrophysics and Cosmology,
Institute for Computational Science,
University of Z\"urich, Winterthurerstrasse 190,
CH-8057, Z\"urich, Switzerland}
\affiliation{Physics Institute, University of Z\"urich,
Winterthurerstrasse 190, CH-8057, Z\"urich, Switzerland}

\author{Ermis Mitsou}
\author{Yves Dirian}
\affiliation{Center for Theoretical Astrophysics and Cosmology,
Institute for Computational Science,
University of Z\"urich, Winterthurerstrasse 190,
CH-8057, Z\"urich, Switzerland}

\author{Ruth Durrer}
\affiliation{D\'epartement de Physique Th{\'e}orique \& Center
for Astroparticle Physics, Universit\'e de Gen\`eve\\
Quai E. Ansermet 24, CH-1211 Gen\`eve 4, Switzerland}

\date{\today}

\begin{abstract}
The background photon temperature~$\Tbar$ is one of the fundamental 
cosmological parameters, and it is often set equal
to the precise measurement $\Tbobs$ of the comic 
microwave background (CMB) temperature by COBE FIRAS.
However, even in future CMB experiments,
$\Tbar$~will remain unknown due to the unknown monopole 
contribution~$\Theta_0$ at our position to the observed (angle-averaged) 
temperature~$\Tbobs$. Using the Fisher formalism, we find that
the standard analysis with $\Tbar\equiv\Tbobs$  underestimates the error
bars on cosmological parameters by $1\sim2\%$ of the present errors,
and the best-fit parameters obtained in the analysis are biased by 
$\sim$1\% of their standard deviation.
These systematic errors are negligible for the {\it Planck} 
data analysis, providing a justification to the standard practice.
However, with~$\Tbar\equiv\Tbobs$,
these systematic errors will always be 
present and irreducible, and future cosmological surveys
 might misinterpret the measurements.
\end{abstract}

\maketitle

{\it Introduction.---}
 Cosmology has seen enormous development in recent
decades (see, e.g., \cite{WEMOET13} for a review). In particular, the
cosmic microwave background (CMB) experiments have greatly improved 
in recent years with the Wilkinson Microwave Anisotropy Probe (WMAP)
and the {\it Planck} satellites \cite{SPVEET03,PLANCK13}.
The primary cosmological parameters are now
constrained at the sub-percent level \cite{PLANCKcos18,PLANCKover18}, 
and the angular scale of
the acoustic peak is even better constrained by an order-of-magnitude.
This level of precision in cosmological parameter estimation demands
a matching accuracy in our theoretical predictions.

The background CMB
temperature~$\Tbar$ is one of the fundamental cosmological parameters
that characterize the evolution of the Universe. In particular, it is 
tantamount to the photon energy density~$\omega_\ga$, and it sets 
the total radiation density~$\omega_r$ (hence the epoch~$z_\up{eq}$ of the
matter-radiation equality), once the other cosmological parameters such as
the matter density~$\omega_m$ and the neutrino masses $m_\nu$ are provided.
Despite its significant role in cosmology, the background CMB 
temperature~$\Tbar$ has rarely been treated as a free cosmological parameter
in literature, because of the pioneering work 
\cite{FIRAS94,FICHET96,FIXSE09}
by the COBE Far Infrared Absolute Spectrometer (FIRAS) in 1990, which
provided the precise measurements of the observed CMB 
temperature~$\Tbobs$ at our position
by averaging the CMB temperature measurements
over the sky. 

The final release \cite{FICHET96}
of the COBE FIRAS measurements is $\Tbobs=2.728\pm0.004$~K,
and the measurements were later further calibrated in Ref.~\cite{FIXSE09}
by using the WMAP differential temperature measurements \cite{WMAP09}:
$\Tbobs=2.7255\pm5.7\cdot10^{-4}$~K. This measurement of the CMB temperature
with exquisite precision underpins the standard practice, in which the
background CMB temperature~$\Tbar$ is set equal to the observed CMB
temperature~$\Tbobs$  {\it without any error} associated to this number.
Ref.~\cite{HAWO08} investigated the impact of the measurement error of 
$\Tbobs$ on the other cosmological parameters and found a negligible inflation 
of their error bars.

In this {\it Letter}, we show that this practice 
is {\it formally incorrect}, 
because it neglects the uncertainty related to cosmic variance 
\cite{MIYOET19}, i.e. the fact that we can only observe a single light-cone.
 Instead, $\Tbar$ should {\it in principle} be considered as an extra free
 cosmological parameter to be varied in the Bayesian analysis.
With $\Tbar\equiv\Tbobs$, the standard practice leads to underestimation
of the error bars on the cosmological parameters
(consistent with the results in \cite{HAWO08}), and systematic biases
in the cosmological parameter estimation (an effect absent in \cite{HAWO08}),
even in the era of future CMB experiments with virtually {\it no}
measurement errors in~$\Tbobs$.
Although the overall impact on parameter estimation is negligible today, 
it might become relevant in the future.

{\it The cosmological parameter~$\Tbar$.---} The background CMB 
temperature~$\Tbar$ is really one of the other cosmological parameters
such as the background matter density~$\omega_m$ or the (background) Hubble
parameter~$H_0$ that are defined in a homogeneous and isotropic universe
and control the evolution of the perturbations in an inhomogeneous universe.
The observed CMB temperature~$\Tbobs$ from the COBE FIRAS is, on the other
hand, obtained
by averaging the CMB temperature measurements on the sky, and it
differs from the background CMB temperature~$\Tbar$ due to the monopole
perturbation~$\Theta_0$. As any other physical quantities, the CMB temperature
 at a given position~$x$ and direction~$\nhat$ in general 
includes not only the background~$\Tbar$, but also the 
perturbation~$\Theta(x,\nhat)$, and
the separation of the background and perturbation is made
for our theoretical convenience. Therefore, when averaged over the sky
at our position~$x_o$, the observed CMB temperature can be expressed as
$\Tbobs=\Tbar(1+\Theta_0)$, where the monopole perturbation is
\beeq
\Theta_0:=\int{d^2\nhat\over4\pi}~\Theta(x_o,\nhat)~,
\eneq
and we suppressed the dependence of~$\Theta_0$ on the observer position~$x_o$.

Compared to the other multipole moments~$\Theta_l$ ($l\geq1$) in CMB,
the monopole is {\it not} an observable, as it is absorbed into the observed
CMB temperature~$\Tbobs$ together with the background temperature~$\Tbar$. 
Despite this peculiarity, 
the monopole perturbation~$\Theta_0$ at our position is very unlikely to be
zero. The Ergodic theorem states that once the fluctuations are averaged over
a sufficiently large volume, the resulting average is equivalent to the 
ensemble average or the average over many realizations of our Universe.
While the ensemble average of the monopole is zero, it is shown 
in Ref.~\cite{MIYOET19} that the angle average
is {\it not} quite the ensemble average, as it is obtained only at our own 
position. This implies that if we were to perform the angle average of
the CMB temperature at the Andromeda galaxy, we would obtain~$\Tbobs$
different from the COBE FIRAS result, due to the fluctuation of the monopole
from place to place. Only if we could average the CMB 
temperature~$\Tbobs(x)$ over all the possible observer positions,
we would be able to replace the average with the ensemble average and
obtain the background CMB temperature~$\Tbar$. 
As this procedure is impossible,
the background CMB temperature~$\Tbar$ can {\it never} be measured and 
needs to be treated as a free
cosmological parameter as the other cosmological parameters.

As an extra cosmological parameter in the Bayesian analysis, the prior 
distribution of~$\Tbar$ should have a mean of $\Tbobs$ and a standard deviation
$\sigma_{\ln\Tbar} \simeq (\sigma_{\Theta_0}^2 + \sigma_m^2)^{1/2}$,
where $\sigma_m\sim2\cdot10^{-4}$ is the current measurement uncertainty
and $\sigma_{\Theta_0}\sim 10^{-5}$ is the cosmic variance contribution 
of the monopole. Since currently $\sigma_m \sim 20 \sigma_{\Theta_0}$,
the effect of cosmic variance will be negligible as well. However, the fact 
that $\sigma_m$ is already close to $\sigma_{\Theta_0}$ implies that future
CMB measurements might cross the threshold.
Note that the {\it Planck} team did allow $\bar{T}$ to vary in their 
analysis \cite{PLANCKcos15}, 
but by {\it ignoring} the COBE FIRAS input at the prior level. The aim of 
this exercise was to establish how well $\bar{T}$ can be constrained 
by the anisotropy and galaxy clustering data {\it alone} and whether the 
result would be consistent with the COBE FIRAS measurement of $\Tbobs$,
under the assumption $\Tbar\equiv\Tbobs$.

{\it CMB observations and theoretical predictions.---} In observations, 
the CMB temperature map as well as the 
polarization map obtained in the CMB experiments is decomposed with spherical
harmonics~$Y_{lm}$ as
$\Tobs(\nhat):=\sum_{lm}\Tobs_{lm}Y_{lm}(\nhat)$, 
and the angular multipoles~$T_{lm}$
are used to construct the observed CMB power spectra 
$D^\up{obs}_l:=\sum_m|\Tobs_{lm}|^2/(2l+1)$
for $l\geq1$. The angle average of the CMB temperature is equivalent
to the monopole $\Tbobs\equiv\Tobs_{00}/\sqrt{4\pi}$.
The theoretical predictions are, however, based on the 
separation of the background and the perturbation around it, so that the CMB
temperature is modeled as $\Tobs(\nhat):=\Tbar(1+\Theta)$ and the angular
decomposition of the temperature anisotropies
$\Theta(\nhat):=\sum_{lm}a_{lm}Y_{lm}(\nhat)$ yields
the angular multipole~$a_{lm}$ and their power spectra~$C_l:=\left\langle
|a_{lm}|^2\right\rangle$, where the angular multipoles and the power spectra
are both dimensionless, as opposed to the dimensionful 
quantities~$\Tobs_{lm}$ and~$D_l^\up{obs}$ in observation.

The conversion between these quantities  is trivial
in theory: $T_{lm}\equiv\Tbar a_{lm}$ and $D_l\equiv\Tbar^2C_l$ for $l\geq1$, 
but 
it is impossible in observation, as the background CMB temperature~$\Tbar$ is 
unknown. However, this poses {\it no} problem, as we can include
an additional cosmological parameter~$\Tbar$
in our data analysis and obtain the best-fit value for~$\Tbar$ as the
other (unknown) cosmological parameters in a given model. The problems arise
because the data analysis is performed by fixing $\Tbar\equiv\Tbobs$ by hand.
This procedure results in two problems: 1)~the background evolution
in our theoretical predictions never matches the correct background
in our Universe, unless the  monopole
at our position happens to be zero, and
2)~by using~$\Tbobs$ instead of~$\Tbar$,
the observed temperature and the CMB power spectra are in practice compared to 
$\Tobs_{lm}/\Tbobs=a_{lm}/(1+\Theta_0)$ and
\beeq
C_l^\up{biased}:=\left\langle{|a_{lm}|^2\over(1+\Theta_0)^2}\right\rangle
=C_l\left(1+{3\over4\pi}C_0+\cdots\right)~,
\eneq
where the monopole of the power spectrum is $C_0\simeq1.7\cdot 10^{-9}$ in our 
fiducial $\Lambda$CDM model. Though negligible in the {\it Planck} data 
analysis, the point~1)
causes systematic errors in the standard data analysis
larger than the point~2).

{\it Underestimation of the error bars.---} One immediate consequence
of the standard practice with $\Tbar\equiv\Tbobs$ is the underestimation
of the error bars on the cosmological parameters in a given model, as there
exists one less degree of freedom in the parameter estimation than 
in reality. The true error bars on the cosmological parameters can be
estimated by considering the full model with extra cosmological 
parameter~$p_0:=\ln\Tbar$, in addition to the standard model parameters~$p_i$
($i=1,\cdots,N$) and by marginalizing over the nuisance parameter~$p_0$.
To estimate the inflation of the error bars, we adopt the Fisher information
matrix formalism.
For the Gaussian fluctuations on the sky, the Fisher matrix takes
the standard form with one critical difference: the observables contain
both the background and the perturbation. For CMB, the
observables are $\Tobs_{lm}$ and $D^\up{obs}_l$, and the Fisher matrix is 
then obtained in Ref.~\cite{YOMIET19} as
\bear
F_{00}&=&{4\pi\over C_0}+\sum_{l=2}^\infty
{2l+1\over2~C_l^2}\left(2C_l+{\pa ~ C_l\over\pa\ln\bar T}\right)^2~,\\
F_{i0}&=&\sum_{l=2}^\infty{2l+1\over2~C_l^2}\left({\pa\over\pa p_i}C_l\right)
\left(2C_l+{\pa~ C_l\over\pa\ln\bar T}\right)~,\\
F_{ij}&=&\sum_{l=2}^\infty
{2l+1\over2~C_l^2}\left({\pa\over\pa p_i} C_l\right)
\left({\pa\over\pa p_j} C_l\right)~,
\enar
where the standard Fisher analysis corresponds to
the sub-matrix of the full Fisher matrix ($F^\up{std}_{ij}\equiv F_{ij}$).
The true error bars on the cosmological parameters after marginalizing
over~$p_0$ can be obtained as the diagonal elements of the $N$-$N$ sub-matrix
\beeq
\sigma^2_p=\up{diag.}
\left(F_{ij}-{F_{i0}F_{0j}\over F_{00}}\right)^{-1}~
\eneq
of the inverse of the full Fisher information matrix. 

For the proof of concept, we apply the Fisher formalism to a CMB experiment
like the {\it Planck} satellite, where we used the temperature $C_l^\up{TT}$
at $l=2\sim2500$, the polarization $C_l^\up{EE}$ at $l=2\sim2000$,
 and the cross $C_l^\up{TE}$ power spectra at $l=30\sim2000$ 
as our CMB observables. The Fisher
matrix is computed  by accounting for the covariance among the temperature
and the polarization observables \cite{ZASE97,ZASPSE97}.
We adopt that the sky coverage is
$f_\up{sky}=0.86$, the detector pixel noise is 
$\Delta^2_T=(0.55\mu\rm K~ deg)^2$, and
the beam size is $\sigma_b=7.22$~arcmin in FWHM for 143~GHz channel. These
specifications are taken into consideration in the Fisher matrix by
modifying the factor~$(2l+1)/2C_l^2$. Finally,
for our fiducial cosmological parameters, we adopted the best-fit $\Lambda$CDM
model parameters reported in Table~7 of the {\it Planck} 2018 results 
\cite{PLANCKover18} ({\it Planck} alone). The CMB power spectra are
computed by using the {\footnotesize CLASS} Boltzmann code \cite{CLASS}.

\begin{figure}[t]
\centering
\includegraphics[width=0.5\textwidth]{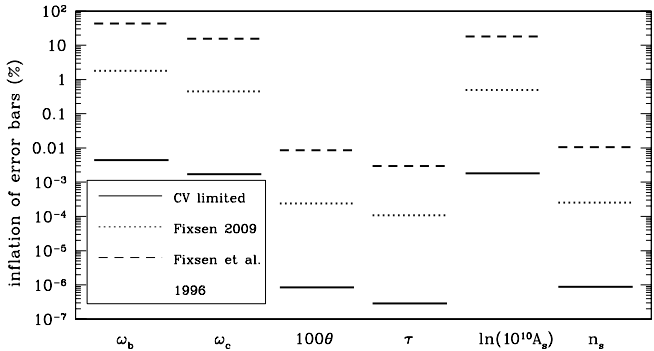}
\caption{Inflation of the error bars on the $\Lambda$CDM cosmological
parameters, after the unknown background temperature~$\Tbar$ is accounted 
for. The errors are relative, e.g., 1\% in the plot means that the true
error bar~$\sigma$ is larger than~$\sigma_\up{std}$ in the standard practice
by 1\%: $\sigma=1.01\sigma_\up{std}$.
By fixing $\Tbar\equiv\Tbobs$, the error bars on the
cosmological parameters are {\it underestimated} in the standard data
analysis. Solid lines represent
the future CMB experiment, in which {\it no} measurement errors exist
in the observed CMB temperature~$\Tbobs$ and only the cosmic
variance contributes to the difference between~$\Tbar$ and~$\Tbobs$.
Dotted lines show the current status, in which the temperature measurement 
by FIRAS was calibrated with the WMAP data \cite{FIXSE09}:
$\Tbobs=2.7255\pm5.7\cdot10^{-4}$~K. Dashed lines show the previous status,
representing the original FIRAS temperature measurement \cite{FICHET96}:
$\Tbobs=2.728\pm0.004$~K.}
\label{fig:f1}
\end{figure}

Figure~\ref{fig:f1} illustrates the underestimation of the true error bars
on the cosmological parameters in the standard practice. We consider 
three cases, in which the observed CMB temperature~$\Tbobs$ is constrained
with different precision: no measurement uncertainty ($\sigma_m\equiv0$; 
solid), COBE FIRAS 
measurement uncertainty calibrated with the WMAP measurements (dotted),
and original COBE FIRAS measurement uncertainty (dashed). In {\it none} 
of these three cases,
we have the precise information about the background CMB temperature~$\Tbar$.
However, given the monopole power  $C_0\simeq1.7\cdot10^{-9}$,
the 1-$\sigma$ rms fluctuation of the monopole is
$\Theta_0\equiv a_{00}/\sqrt{4\pi}\sim1.2\cdot10^{-5}$, so the 
background CMB temperature~$\Tbar$ is likely to be within the current
measurement uncertainty $6\cdot10^{-4}$~K from~$\Tbobs=2.7255$~K.

Under the assumption that the monopole happens to vanish at our
position, the  standard data analysis underestimates the
error bars on the cosmological parameters, for instance, 
by {\it two percent} 
for the baryon density~$\omega_b$, when the measurement 
of~$\Tbobs$ from COBE FIRAS is calibrated with the WMAP measurements 
and by {\it tens of percents} 
when the original COBE FIRAS measurement is used.
Note that the inflation of error bars in Figure~\ref{fig:f1} is relative
to the error bar in the standard practice.
The amplitude~$A_s$ of the curvature perturbation is equally affected,
while the angular size~$\theta$ or the spectral index~$n_s$ are less
sensitive. The inflation of the error bars is largely determined by 
two factors: the uncertainty in~$\Tbar$ (or $C_0$ in~$F_{00}$)
and the correlation~$F_{i0}$ of the parameter~$p_i$ and the temperature~$\Tbar$
variations. $F_{i0}$ is stronger for~$\omega_b$ and~$\omega_c$, and this
trend is amplified by the correlation $F^{-1}_\up{std}$ among the model
parameters. The error bars in~$A_s$ is enhanced largely by the parameter
correlation. With an order-of-magnitude reduction of the uncertainty
in~$\Tbobs$ in Ref.~\cite{FIXSE09}, the inflation of the error bars
(dotted) is less than a few percents
for the $\Lambda$CDM cosmological parameters.
 Propagating the errors on~$\omega_b$, $\omega_c$, and $100\theta$,
we obtain the inflation of the error on the Hubble parameter~$h$: 2\%, 0.04\%,
$10^{-4}\%$ for the three cases.
What is important is to note that the error bars are {\it always}
underestimated (solid lines) in the standard data analysis, even 
with {\it no} measurement uncertainty in~$\Tbobs$ from future CMB 
experiments.

{\it Cosmological parameter bias.---} By fixing $\Tbar\equiv\Tbobs$,
the standard data analysis contains systematic errors in terms of
biases in the cosmological parameter estimation. Assuming that the systematic
errors are small, the best-fit cosmological parameters~$p^b_\mu$
are characterized by the parameter biases~$\de p_\mu$ from the true parameter 
set~$p^t_\mu$ as $p^b_\mu:=p^t_\mu+\de p_\mu$ ($\mu=0,1,\cdots,N$), where 
in the standard practice 
$p^b_0\equiv\ln\Tbobs=\ln[\Tbar(1+\Theta_0)]\simeq\ln\Tbar+\Theta_0$, so that
the parameter bias for $p_0=\ln\Tbar$ is the unknown monopole at our position:
$\de p_0\equiv\Theta_0$.

The relation between two parameter sets can be obtained by considering that
the likelihood~${\cal L}(p_\mu)$ of the CMB observables is maximized at the 
best-fit parameters~$p^b_\mu$:
\bear
0&=&{\pa\over\pa p_i}{\cal L}\bigg|_{p^b_\mu}=\Tr\left[\tilde{\bdv{C}}^{-1}
\tilde{\bdv{C}}_{,i}\right] \\
&&
-\Tr\left[\tilde{\bdv{C}}^{-1}\tilde{\bdv{C}}_{,i}\tilde{\bdv{C}}
^{-1}\left(\Dobs-\tilde\mu\right)\left(\Dobs-\tilde\mu\right)^\up{T}
\right]~,\nn
\enar
where the commas represent derivative of the covariance matrix~$\bdv{C}$
with
respect to the parameter~$p_i$ and the observed data set~$\Dobs$ includes
the observed temperature and polarization anisotropies. The covariance
matrix~$\bdv{C}(p_\mu)$ and the mean~$\mu(p_\mu)$ 
are the theoretical predictions in a given model, where $\mu=\Tbar$ for
temperature anisotropies and $\mu=0$ for polarization anisotropies.
However, due to the  assumption $\Tbar\equiv\Tbobs$
in the standard practice,
the theoretical predictions for~$\bdv{C}$ and~$\mu$ depend only on the
model parameters~$p_i$, but {\it not} on~$\Tbar$, 
and we used tilde to represent that the theoretical predictions are 
evaluated at~$p^b_\mu$, not at~$p^t_\mu$.

Using the spherical harmonics decomposition, the condition for the best-fit 
parameter set is expressed as
\beeq
0=\sum_{l=2}^\infty(2l+1)\tilde C_l^{-1}{\pa\over\pa p_i}\tilde C_l
\left[1-{1\over2l+1}\sum_m{\Tbar^2|\aobs_{lm}|^2
\over(\Tbobs)^2\tilde C_l}\right]~,
\eneq
where the power spectra~$\tilde C_l$ account for the covariance among the
temperature, the polarization, and their cross power spectra
together with the detector noise and beam smoothing \cite{ZASE97,ZASPSE97}.
To make further progress, we take the ensemble average to replace the
ratio of~$\aobs_{lm}$ and~$\Tbobs$ 
with~$C_l^\up{biased}$ and expand the power spectra around~$p^b_\mu$ as
\beeq
C_l^\up{biased}(p^t_\mu)\simeq \tilde C_l
\left(1+{3\over4\pi}\tilde C_0-{\pa\ln\tilde C_l\over\pa\ln\Tbar}\Theta_0
-{\pa\ln\tilde C_l\over\pa p_i}\de p_i\right)~,
\eneq
where the first correction arises from~$C_l^\up{biased}$ and the remaining
corrections arise due to the difference between~$p^b_\mu$ and~$p^t_\mu$.
Ignoring the small correction due to the first term, 
the cosmological parameter bias can be neatly expressed as
\beeq
\de p_i=-\left(F_\up{std}^{-1}\right)_{ij}F_{j0}~\Theta_0~,
\eneq
and it is in proportion to the amplitude of the unknown
monopole at our position, while it is independent of the measurement
uncertainty in~$\Tbobs$, given our assumption $p^t_\mu\simeq p^b_\mu$.

\begin{figure}[t]
\centering
\includegraphics[width=0.5\textwidth]{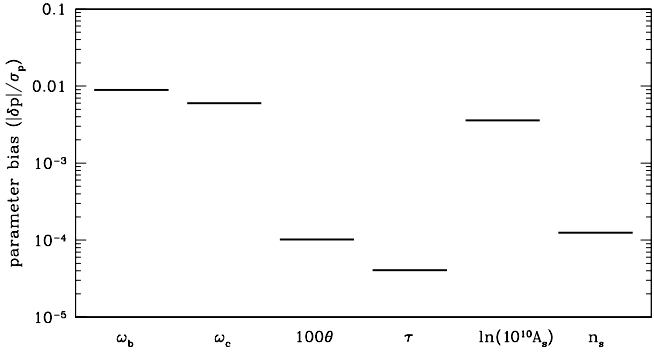}
\caption{Bias $\de p_i$ in the best-fit cosmological parameters,
in terms of the standard deviation~$\sigma_{p_i}$. The amplitude
of the monopole at our position is assumed to be at 1-$\sigma$ fluctuation:
$\Theta_0\equiv1.2\cdot10^{-5}$. The cosmological parameter bias is 
independent of the measurement uncertainty in~$\Tbobs$, but in proportion
to the amplitude of the monopole.}
\label{fig:f2}
\end{figure}

Figure~\ref{fig:f2} shows the bias~$\de p_i$ in units of the parameter's
standard deviation~$\sigma_{p_i}$ in the best-fit cosmological
parameters with $\Theta_0$ assumed to be at 1-$\sigma$ fluctuation.
If the monopole happens to {\it vanish} at our position, there would be 
{\it no} bias in the cosmological parameters by using the standard practice. 
However, if the monopole at our position is {\it non-zero}, 
the standard analysis
yields the biases in the best-fit cosmological parameters 
in proportion to the unknown amplitude of the monopole. For instance,
the baryon density parameter~$\omega_b$ is
off by 0.01$\sigma_{\omega_b}$ at 1-$\sigma$ fluctuation of~$\Theta_0$, and
this level of bias is {\it readily tolerable} today. 
While the biases in~$\omega_c$ and~$\ln(10^{10}A_s)$ are of similar magnitude, 
their error bars are larger, hence the impacts are slightly smaller. 
The impacts for $100\theta$,~$\tau$, and~$n_s$ are negligible.

{\it Conclusions.---} We showed that in principle
the background CMB temperature~$\Tbar$ has to be considered as
an unknown cosmological parameter, because
the observed (angle-average) CMB 
temperature~$\Tbobs$ includes the unknown monopole contribution at our 
position. We investigated the impact of this ``new'' cosmological
parameter~$\Tbar$ on the CMB data analysis.
With the current uncertainty in~$\Tbobs$,
the standard data analysis underestimates the
error bars on the cosmological parameters by a relative amount of up to 2\%,
and if the monopole is {\it non-vanishing} at our position, 
the best-fit cosmological parameters in the standard analysis are biased 
by about 1\% of their current standard deviation or 1-$\sigma$ error bar.

We conclude that these systematic errors are {\it negligible} in the
{\it Planck} data analysis, providing a further justification to the standard
practice. However, these systematic errors
are {\it always} present and irreducible in the standard data analysis,
so that cosmological measurements might be misinterpreted
in future experiments with better precision than the {\it Planck} satellite.
Of course, these systematic errors can be readily avoided 
by including one extra cosmological parameter~$\Tbar$. 

We thank Antony Lewis,  Pavel Motloch, Douglas Scott, David Spergel, 
Matias Zaldarriaga, and James Zibin for useful discussions. 
We acknowledge support
by the Swiss National Science Foundation. J.Y., E.M., Y.D.
are further supported by
a Consolidator Grant of the European Research Council (ERC-2015-CoG grant 
680886).

\bibliography{ms.bbl}

\end{document}